\documentclass[3p,times,twocolumn]{elsarticle}

\usepackage{ecrc}

\volume{00}

\firstpage{1}

\journalname{Nuclear Physics B Proceedings Supplement}

\runauth{}

\jid{nuphbp}

\jnltitlelogo{Nuclear Physics B Proceedings Supplement}

\usepackage{amssymb}

\usepackage[figuresright]{rotating}

\begin{document}

\begin{frontmatter}


\author{Soebur Razzaque\corref{cor1}\fnref{label1}}
\ead{srazzaque@uj.ac.za}
\author{A. Yu. Smirnov\corref{cor2}\fnref{label2,label3}}
\ead{smirnov@mpi-hd.mpg.de}

\dochead{}

\title{Super-PINGU for measuring CP violation}

\address[label1]{Department of Physics, University of Johannesburg, PO Box 524, Auckland Park 2006, South Africa}
\address[label2]{Max-Planck-Institute for Nuclear Physics, Saupfercheckweg 1, D-69117 Heidelberg, Germany}
\address[label3]{International Centre for Theoretical Physics, Strada Costiera 11, I-34100 Trieste, Italy}



\begin{abstract}
We propose to measure leptonic CP phase, after neutrino mass hierarchy is established, with an upgrade of the PINGU detector and using atmospheric neutrino flux. The upgrade, called super-PINGU, will require a few megaton effective volume at 0.5-1 GeV range to distinguish $\delta$ in the range $\pi/2$-$3\pi/2$ from 0 after 4 years of operation. The distinguishability (similar to significance) of measuring $\delta$ depends crucially on various flux, cross-section, event reconstruction (energy and angle) and flavor identification uncertainties. We explore effects of these uncertainties on the distinguishability of measuring CP phase and suggest possible ways to minimize their impact. 
\end{abstract}


\end{frontmatter}


\section{Introduction}
\label{intro}

Establishing CP violation in the leptonic sector is an outstanding problem in particle physics. Atmospheric neutrino flux measurements in large water/ice detectors can be used to determine the Dirac CP phase $\delta$. Information on different $\delta$ values is encoded in the neutrino oscillation probabilities after propagation inside the earth (matter effect). A systematic shift of the probabilities with increasing $\delta$ in a wide, $\sim 0.2$--2 GeV, energy range is key to measure $\delta$ with atmospheric neutrino flux \cite{Razzaque:2014vba}.

It has been found recently that the PINGU and ORCA detector with $\sim 3$ GeV threshold will have good sensitivity to determining the neutrino mass hierarchy \cite{Akhmedov:2012ah, Ribordy:2013xea, Aartsen:2014oha, Katz:2014tta}. However, measurement of $\delta$ would require a detector with larger effective volume and improved characteristics in the $< 2$ GeV range and in this context a future upgrade of PINGU (and also of ORCA), called Super-PINGU, was proposed in Ref.~\cite{Razzaque:2014vba} with detailed estimate of sensitivity. Here we present highlights from that work.

\section{Methodology and Results}
\label{results}

To measure CP phase with atmospheric neutrinos, we calculate neutrino events in Super-PINGU by varying $\delta$ and compare with $\delta = 0$, keeping all other oscillation parameters fixed. We use an effective mass (both for $\nu_\mu$ and $\nu_e$) parameterized as
\begin{equation}
\rho V_{\rm eff} (E_\nu) = 2.6 \left[\log (E_\nu/{\rm GeV}) + 1 \right]^{1.32} ~{\rm Mt},
\end{equation} 
which can be realized for a total of 126 strings and 60 DOMs per string. This gives an effective mass of $\sim 2.8$ Mt at $\sim 1$-2 GeV, which is 4 times larger than PINGU. The number of neutrino events for a particular flavor $\alpha = e, \mu$ with energies and zenith angles in small bins  $\Delta (E_\nu)$ and $\Delta(\cos \theta_z)$ marked by subscript  $ij$ can be calculated as
\begin{eqnarray} 
N_{ij, \alpha} &=& 2 \pi N_A \rho T 
\int_{\Delta_i\cos\theta_z} d\cos\theta_z \cr
&& \int_{\Delta_jE_\nu} dE_\nu~ 
V_{\rm eff, \alpha} (E_\nu) d_\alpha (E_\nu, \theta_z).
\label{eq:nev} 
\end{eqnarray}
Here $T$ is the exposure time, $N_A$ is the Avogadro's number. The density of events of type $\alpha$, $d_\alpha$ (the number of events per unit time per target nucleon), is  given by $d_\alpha(E_\nu, \theta_z)  = \sigma_\alpha \Phi_\alpha  + {\bar \sigma}_\alpha {\bar \Phi}_\alpha$ in terms of fluxes at the detector, and $\Phi_\alpha =  \Phi_\mu^0 P_{\mu \alpha} + \Phi_e^0 P_{e \alpha}$ with corresponding oscillation probabilities $P_{\mu \alpha} $ and $P_{e \alpha}$ . The original muon and electron neutrino fluxes at the production are $\Phi_\mu^0 = \Phi_\mu^0 (E_\nu,\theta_z) $ and $\Phi_e^0 = \Phi_e^0(E_\nu,\theta_z)$.

\begin{figure}[t]
\includegraphics[width=6.15cm]{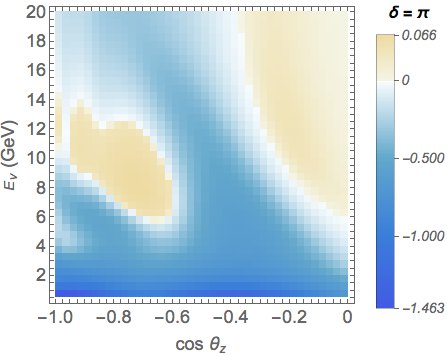}
\includegraphics[width=6.15cm]{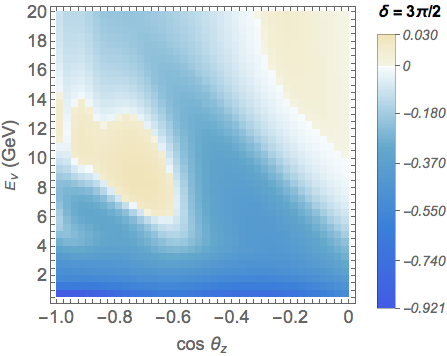}
\caption{Distributions of $S_{ij}$ for $\nu_\mu + {\bar \nu}_\mu$ events in case of different 
$\delta$ values and $\delta = 0$ in the $E_\nu$--$\cos\theta_z$ plane. $E_\nu$ and $\theta_z$ are reconstructed neutrino energy and direction after smearing of the true energy and direction using PINGU reconstruction functions with a factor  $\sqrt{3}$ better resolutions. Normal mass hierarchy is assumed.}
\label{fig1}
\end{figure}

\begin{figure}[t]
\includegraphics[width=5.95cm]{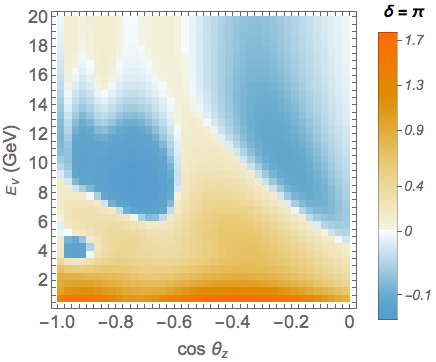}
\includegraphics[width=5.95cm]{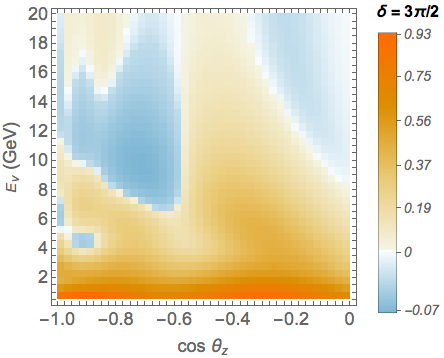}
\caption{The same as Fig.~\ref{fig1} but for $\nu_e + {\bar \nu}_e$ events.}
\label{fig2}
\end{figure}

We have computed the  distributions of $\nu_\mu + {\bar \nu}_\mu$ and $\nu_e + {\bar \nu}_e$ events for $\delta = 0$ and $\delta \ne 0$ and take difference between the distributions in the $E_\nu$--$\cos \theta_z$ plane to study their properties. Since there are errors associated with reconstructing the true neutrino energy and directions, we smear the ideal distributions with the energy and angular resolution functions of the detector to mimic the real situation. To estimate the sensitivity of measuring a CP phase   different from zero we employ a distinguishability parameter defined as
\begin{equation}
S_{ij} (f) = (N_{ij}^\delta - N_{ij}^0)/\sigma_{ij} (f),
\end{equation}
where $N_{ij}^\delta$ and $N_{ij}^0$ are the reconstructed number of events in the $ij$-th bin in the $E_\nu$--$\cos \theta_z$ plane for $\delta$ and $\delta = 0$, respectively, and $\sigma_{ij}^2 (f) = N_{ij}^0 + (f N_{ij}^0)^2$ is the total error in the $ij$-th bin. Parameter $f$ is a measure of uncorrelated systematic errors \cite{Akhmedov:2012ah} and we take $f = 2.5\%$. The total distinguishability
\begin{equation}
S_\sigma = \sqrt{\sum_{ij} S_{ij}^2}
\end{equation}
is a quick measure of significance. 

Figure \ref{fig1} shows the distinguishability $S_{ij}$ for $\nu_\mu + {\bar \nu}_\mu$ events with $f=0$ for different $\delta$ values and using 1-year of Super-PINGU data. Normal mass hierarchy is assumed. The shape of the distributions, specially their domain structures, is largely explained as due to grids of solar, atmospheric and interference magic lines in the $E_\nu$--$\cos\theta_z$ plane. The oscillation probabilities are independent of $\delta$ along these lines, thus separating regions of same sign distinguishability. Figure \ref{fig2} shows $S_{ij}$ distributions for $\nu_e + {\bar \nu}_e$ events.

\begin{figure}[t]
\includegraphics[width=7.8cm]{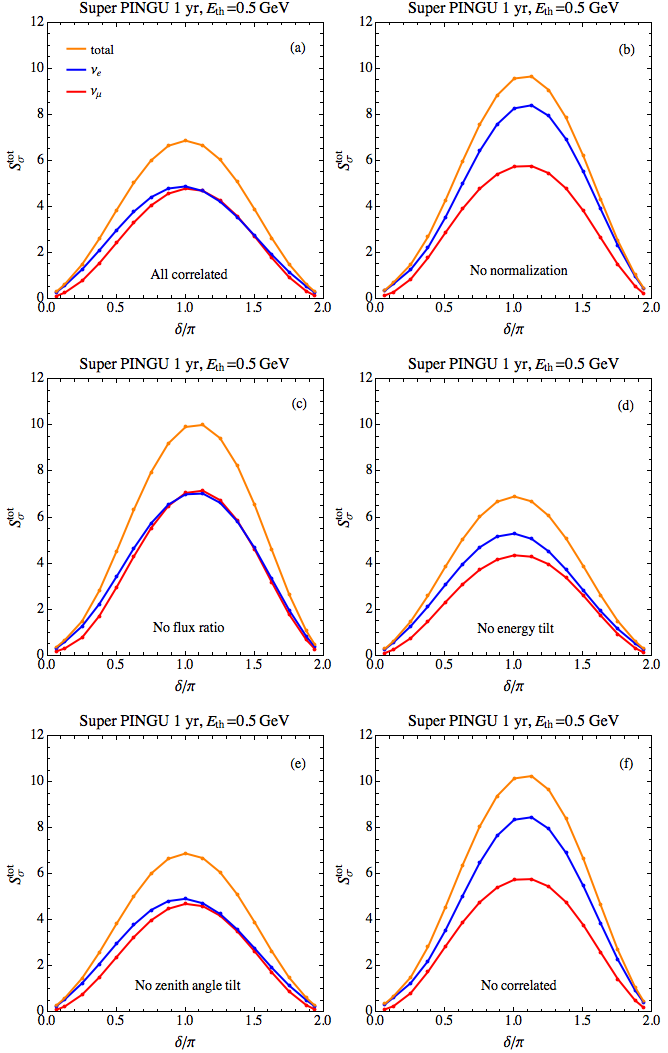}
\caption{Distinguishability of measuring CP phase in the $\nu_\mu$ and $\nu_e$ channels, with effects of various correlated uncertainties. $2.5\%$ uncorrelated uncertainty has been assumed in all cases.}
\label{fig3}
\end{figure}

The uncertainties associated with atmospheric neutrino flux, $\nu N$ cross section, effective volume, etc.\ affect neutrino event distributions across bins in the $E_\nu$--$\cos\theta_z$ plane. We include effects of these correlated uncertainties in our calculation with analogy to the pull method in $\chi^2$ analysis. In particular we minimize the following distinguishability parameter
\begin{eqnarray}
S_\sigma^{tot} (\xi_k) &=& \left[ 
\sum_{l=e,\mu} \sum_{ij} \frac{[N_{ij,l}^\delta (\xi_k) - N_{ij,l}^0 (\xi_k^{st})]^2}{\sigma_{ij,l}^2} \right. \cr
&& \left. + \sum_k \frac{(\xi_k - \xi_k^{st})^2}{\sigma_k^2} \right]^{1/2},
\end{eqnarray}
where $\xi_k$ are the pull variables and $\xi_k^{st}$ are their standard values. The event distributions with varying $\xi_k$ are calculated as
\begin{eqnarray}
N_{ij,l}^\delta (\xi_k) &=& \alpha z_l ( E/2~{\rm GeV})^\eta \cr
&& \times [ 1+ \beta ( 0.5 + \cos\theta_z ) ] N_{ij,l}^\delta (\xi_k^{st}) ,
\end{eqnarray}
where $\alpha$ is the overall normalization factor  with the error $\sigma_\alpha = 0.2$, $z_l$ is the flux (flavor) ratio  uncertainty ($z_e \equiv 1$ for $\nu_e$ events), with  the error $\sigma_z = 0.05$;  $\eta$ is the energy tilt parameter with $\sigma_\eta = 0.1$; $\beta$ is the zenith angle  tilt with  $\sigma_\beta = 0.04$. Figure \ref{fig3} shows the $S_\sigma^{tot}$ minimized over $(\xi_k)$ for different correlated uncertainties as well as for no correlated uncertainties. A threshold energy of 0.5 GeV has been assumed. Note that the contributions of $\nu_e$ and $\nu_\mu$ channels to $S_\sigma^{tot}$ are comparable. 

\section{Discussion}
\label{discussion}

We estimate that after 4 years of operation and $2.5\%$ systematics, Super-PINGU with 0.5 GeV threshold will be able to distinguish $\delta = \pi/4, \pi/2, \pi, 3\pi/2$ from zero with $S_{\sigma}^{tot} (\pi/4) = (1 - 3)$,  $S_{\sigma}^{tot} (\pi/2) = (3 - 8)$, $S_{\sigma}^{tot} (\pi) = (6 - 14) $, $S_{\sigma}^{tot} (3\pi/2) = (3 - 8)$. The ranges depend on effects of different correlated systematics. These values are a factor 4--6 improvement over the sensitivity of PINGU to $\delta$ with 3 GeV threshold.

The sensitivity of Super-PINGU to $\delta$ can be further improved with following possibilities:
\begin{itemize}
\item Decrease of energy threshold to 0.2 GeV from 0.5 GeV with a denser array. This may increase sensitivity by $30\%$.  
\item Stringent kinematical cut can be used to create a high-quality event sample with better reconstruction of the neutrino energy, direction and flavor.
\item An increased exposure time will also increase the sensitivity to CP by a factor $\propto\sqrt{t}$.
\item Improved flavor identification at low energies.
\item Increase the density of DOMs or photocathode coverage.
\item Statistical separation between the neutrino and antineutrino events.
\end{itemize}

Our results show that Super-PINGU can be competitive to other proposals for measuring the leptonic CP phase associated with long base-line (LBL) accelerator experiments.




\nocite{*}
\bibliographystyle{elsarticle-num}
\bibliography{martin}



\end{document}